\documentclass[12pt]{iopart}
\usepackage{iopams}
\begin{document}
\letter{Icosahedral multi-component model sets}
\author{Nicolae Cotfas}
\address{Faculty of Physics, University of Bucharest,
PO Box 76-54, Postal Office 76, Bucharest, Romania,\\ 
E-mail address: ncotfas@yahoo.com,\quad 
http://fpcm5.fizica.unibuc.ro/\,{\Huge $\mbox{}_{\mbox{}_{\tilde{ }}}\;\!$}ncotfas }  
\begin{abstract}
A quasiperiodic packing ${\mathcal Q}$ of interpenetrating copies of ${\mathcal C}$,
most of them only partially occupied, can be defined in terms of the strip projection
method for any icosahedral cluster ${\mathcal C}$. We show that in the case when the
coordinates of the vectors of ${\mathcal C}$ belong to the quadratic field 
$\mathbb{Q}[\sqrt{5}]$ the dimension of the superspace can be reduced, namely,
${\mathcal Q}$ can be re-defined as a multi-component model set by using a
6-dimensional superspace.
\end{abstract}
\maketitle

\section{Introduction}

An icosahedral quasicrystal can be regarded as a quasiperiodic packing of copies of
a well-defined icosahedral atomic cluster. Most of these interpenetrating copies 
are only partially occupied. From a mathematical point of view, an
icosahedral cluster ${\mathcal C}$ 
can be defined as a finite union of orbits of a 3-dimensional
representation of the icosahedral group, and 
there exists an algorithm \cite{C1,C2} which leads from ${\mathcal C}$ 
directly to a pattern ${\mathcal Q}$ which can be regarded as a union of
interpenetrating partially occupied translations of ${\mathcal C}$. 
This algorithm, based on
the strip projection method and group theory, represents an extended
version of the model proposed by Katz \& Duneau \cite{K} and independently by Elser 
\cite{E} for the icosahedral quasicrystals.

The dimension of the superspace used in the definition of ${\mathcal Q}$
is rather large, and the main purpose of this
paper is to present a way to reduce this dimension. It is based on 
the notion of {\em multi-component model set}, an extension of the notion 
of {\em model set}, proposed by Baake and Moody \cite{B}.

\section{Quasiperiodic packings of icosahedral clusters}

It is known that the icosahedral group
$Y=235=\left< a,\; b\ |\ a^5=b^2=(ab)^3=e\right>$
has five irreducible non-equivalent representations and its character table is
\begin{equation}\label{table}
\begin{array}{ccccccc}
 & & 1\; e &  12\; a &  15\; b & 20\; ab & 12\; a^2\\
\Gamma _1 & & 1 & 1 & 1 & 1 & 1 \\
\Gamma _2 & & 3 & \tau & -1 & 0 & \tau ' \\
\Gamma _3 & & 3 & \tau ' & -1 & 0 & \tau \\
\Gamma _4 & & 4 & -1 & 0 & 1 & -1\\
\Gamma _5 & & 5 & 0 & 1 & -1 & 0
\end{array} 
\end{equation}
where $\tau =(1+\sqrt{5})/2$ and $\tau '=(1-\sqrt{5})/2$.

A realization of $\Gamma _2$ in the usual 3-dimensional Euclidean space 
$\mathbb{E}_3=(\mathbb{R}^3,\langle ,\rangle )$ is the
representation $\{ T_g:\mathbb{E}_3\longrightarrow \mathbb{E}_3\ |\ g\in Y\}$
generated by the rotations $T_a,\ T_b :\mathbb{E}_3\longrightarrow \mathbb{E}_3$
\begin{equation}\label{Y} \fl \begin{array}{l}
T_a(\alpha ,\beta  ,\gamma )=
\left(\frac{\tau -1}{2}\alpha -\frac{\tau }{2}\beta  +\frac{1}{2}\gamma ,
\ \frac{\tau }{2}\alpha +\frac{1}{2}\beta  +\frac{\tau -1}{2}\gamma ,
\ -\frac{1}{2}\alpha +\frac{\tau -1}{2}\beta  
+\frac{\tau }{2}\gamma \right)\\[1mm]
T_b(\alpha ,\beta  ,\gamma )=(-\alpha ,-\beta  ,\gamma ).
\end{array} \end{equation}
In the case of this representation there are the trivial orbit 
$Y(0,0,0)=\{ (0,0,0)\}$ of length 1, the orbits
\begin{equation}
Y(\alpha ,\alpha \tau ,0)=\{ T_g(\alpha ,\alpha \tau ,0)\ |\ g\in Y\}\qquad 
{\rm where}\quad \alpha \in (0,\infty )
\end{equation}
of length 12 (vertices of a regular icosahedron), the orbits
\begin{equation}
Y(\alpha ,\alpha ,\alpha )=\{ T_g(\alpha ,\alpha ,\alpha )\ |\ g\in Y\}\qquad 
{\rm where}\quad \alpha \in (0,\infty )
\end{equation}
of length 20 (vertices of a regular dodecahedron), the orbits
\begin{equation}
Y(\alpha ,0,0)=\{ T_g(\alpha ,0,0)\ |\ g\in Y\}\qquad 
{\rm where}\quad \alpha \in (0,\infty )
\end{equation}
of length 30 (vertices of an icosidodecahedron), and all the other orbits are 
of length 60. 

Let ${\mathcal C}$ be a fixed icosahedral cluster containing only orbits of
length 12, 20 and 30. It can be defined as  
\begin{equation}
{\mathcal C}=\bigcup_{x\in S}Yx=\bigcup_{x\in S}\{ T_gx\ |\ g\in Y\}
=\{ T_gx\ |\ g\in Y,\ x\in S\}=YS
\end{equation}
where the set $S$ contains a representative of each orbit.
The entries of the matrices of rotations $T_a,\ T_b$ in the basis 
$\{ (1,0,0),(0,1,0),(0,0,1)\}$  
\begin{equation} 
T_a=\frac{1}{2}\left( \begin{array}{ccc}
\tau -1 & -\tau & 1\\
\tau & 1 & \tau -1\\
-1 & \tau -1 & \tau 
\end{array} \right) \qquad 
T_b=\left( \begin{array}{rrr}
-1 & 0 & 0\\
0 & -1 & 0\\
0 & 0 & 1
\end{array} \right)
\end{equation}
belong to the quadratic field $\mathbb{Q}[\sqrt{5}]=\mathbb{Q}[\tau ]$.
Since $\mathbb{Q}[\tau ]$ is dense in $\mathbb{R}$ we can assume that
\[ \fl 
S\!\subset \!\{ (\alpha ,\alpha \tau ,0)\, |\, 
\alpha \!\in \!\mathbb{Q}[\tau ],\ \alpha \!\!>\!\!0\} \cup
\{ (\alpha ,\alpha ,\alpha )\, |\, 
\alpha \!\in \!\mathbb{Q}[\tau ],\ \alpha \!\!>\!\!0\} \cup
\{ (\alpha ,0,0)\, |\, 
\alpha \!\in \!\mathbb{Q}[\tau ],\ \alpha \!\!>\!\!0\} 
\]
without a significant loss of generality in
the description of atomic clusters. 
Since the orbits of $Y$ of length 12, 20 and 30 are symmetric with respect 
to the origin, the cluster ${\mathcal C}$ has the form
\begin{equation}
{\mathcal C}=\{ e_1,e_2,...,e_k,-e_1,-e_2,...,-e_k\}
\end{equation}
and for each vector $e_i=(e_{i1},e_{i2},e_{i3})$ the coordinates 
$e_{i1},\ e_{i2},\ e_{i3}$ belong to $\mathbb{Q}[\tau ]$.

Let $\varepsilon _1=(1,0,...,0),\ \varepsilon _2=(0,1,0,...,0),\ ...,\
\varepsilon _k=(0,...,0,1)$ be the canonical basis of $\mathbb{E}_k$.
For each $g\in Y,$ there exist
the numbers $s_1^g,\ s_2^g,...,s_k^g\in \{ -1;\ 1\}$ and a permutation of 
the set $\{ 1,2,...,k\}$ denoted also by $g$ such that,
\begin{equation}\label{sp}
T_ge_j=s_{g(j)}^ge_{g(j)}\qquad {\rm for\ all\ }
j\in \{ 1,2,...,k\}. 
\end{equation}
{\bf Theorem 1.} \cite{C1,C2} {\it  
The formula 
\begin{equation}
g\varepsilon _j=s_{g(j)}^g\varepsilon _{g(j)} 
\end{equation}
defines the orthogonal representation
\begin{equation}
 g(x_1,x_2,...,x_k)=(s_1^gx_{g^{-1}(1)},s_2^gx_{g^{-1}(2)},...,
s_k^gx_{g^{-1}(k)})
\end{equation}
of $Y$ in $\mathbb{E}_k.$}\\[3mm]
{\bf Theorem 2.} \cite{C1,C2} {\it The subspace
\begin{equation}
E=\left\{ \ (<u,e_1>,<u,e_2>,...,<u,e_k>)\ |\ \  
                                       u\in \mathbb{E}_3\ \right\}
\end{equation}
of $\mathbb{E}_k$ is $Y$-invariant and the vectors 
\[ \fl v_1=\varrho (e_{11},e_{21},...,e_{k1})\qquad 
   v_2=\varrho (e_{12},e_{22},...,e_{k2})\qquad 
   v_3=\varrho (e_{13},e_{23},...,e_{k3})\] 
where $\varrho =1/\sqrt{(e_{11})^2+(e_{21})^2+...+(e_{k1})^2}$
form an orthonormal basis of $E$.}\\[3mm]
{\bf Theorem 3.} \cite{C1,C2} {\it The subduced representation of $Y$ in $E$
is equivalent with the representation of $Y$ in $\mathbb{E}_3,$ and the isomorphism
of representations
\begin{equation}
{\mathcal I}:\mathbb{E}_3\longrightarrow E\qquad 
{\mathcal I}u=(\varrho <u,e_1>,\varrho <u,e_2>,...,\varrho <u,e_k>)
\end{equation}
with the property 
${\mathcal I}(\alpha ,\beta ,\gamma )
=\alpha v_1+\beta v_2+\gamma v_3$ allows us to identify 
the `physical' space $\mathbb{E}_3$ with the subspace $E$ of $\mathbb{E}_k$.}\\[3mm]
{\bf Theorem 4.} \cite{C1,C2} {\it The matrix of the orthogonal projector
 \ $\pi :\mathbb{E}_k\longrightarrow \mathbb{E}_k$ corresponding to $E$ in the basis
$\{ \varepsilon _1,\varepsilon _2,...,\varepsilon _k\}$  is}
\begin{equation}
 \pi =\varrho ^2\left( \begin{array}{cccc}
\langle e_1,e_1\rangle & \langle e_1,e_2\rangle & ... & \langle e_1,e_k\rangle \\
\langle e_2,e_1\rangle & \langle e_2,e_2\rangle & ... & \langle e_2,e_k\rangle \\
...& ... & ... & ...\\
\langle e_k,e_1\rangle & \langle e_k,e_2\rangle & ... & \langle e_k,e_k\rangle 
\end{array}  \right) .
\end{equation}

Let $\kappa =1/\varrho $, \ $\mathbb{L}=\kappa \mathbb{Z}^k$, \ 
$\mathbb{K}=[0,\kappa ]^k=\{(x_1,x_2,...,x_k)\ |\ 0\leq x_i\leq \kappa \}$, 
and let $K=\pi ^\perp (\mathbb{K})$, where
$\pi ^\perp :\mathbb{E}_k\longrightarrow \mathbb{E}_k$, 
$\pi ^\perp x=x-\pi x$ 
is the orthogonal projector corresponding to the subspace
\begin{equation}
E^\perp =\{ x\in \mathbb{E}_k\ |\ \langle x,y\rangle =0\ {\rm for\ all}\ y\in E\}.
\end{equation}
{\bf Theorem 5.} \cite{C1,C2} {\it The $\mathbb{Z}$-module $\mathbb{L}\subset \mathbb{E}_k$
is $Y$-invariant, \ $\pi (\kappa \varepsilon _i)={\mathcal I}e_i$,
that is, $\pi (\kappa \varepsilon _i)=e_i$ if we take into consideration 
the identification ${\mathcal I}:\mathbb{E}_3\longrightarrow E,$ and}
\begin{equation}
\pi (\mathbb{L})=\mathbb{Z}e_1+\mathbb{Z}e_2+...+\mathbb{Z}e_k.
\end{equation}

The pattern defined by using the strip projection method \cite{K}
\begin{equation}
{\mathcal Q}=\left\{ \left. \pi  x\ \right| \ 
x\in \mathbb{L},\ \pi ^\perp x\in K\right\}
\end{equation}
can be regarded as a union of interpenetrating copies of ${\mathcal C}$, most of 
them only partially occupied. 
For each point $\pi x\in {\mathcal Q}$ the set of all the arithmetic neighbours 
of $\pi x$
\[ \{ \pi y\ |\  y\in \{ x+\kappa \varepsilon _1,...,x+\kappa \varepsilon _k,
x-\kappa \varepsilon _1,...,x-\kappa \varepsilon _k\},  \pi ^\perp y\in K\}\]
is contained in the translated copy 
\[ \{ \pi x +e_1,...,\pi x+e_k,\pi x -e_1,...,\pi x-e_k\}=\pi x +{\mathcal C} \]
of the $G$-cluster ${\mathcal C}$.
The fully occupied clusters occuring in ${\mathcal Q}$ correspond to the points 
$x\in \mathbb{L}$ satisfying the condition \cite{K}
\begin{equation}
\pi ^\perp x\in K\cap \bigcap_{i=1}^k(\pi ^\perp (\kappa \varepsilon _i)+K)\cap 
\bigcap_{i=1}^k(-\pi ^\perp (\kappa \varepsilon _i)+K).
\end{equation}
Generally, only a small part of the clusters occuring in ${\mathcal Q}$ can
be fully occupied.
A fragment of ${\mathcal Q}$ can be 
obtained by using, for example, the algorithm presented in \cite{V}. The main 
difficulty is the rather large dimension $k$ of the superspace $\mathbb{E}_k$
used in the definition of ${\mathcal Q}$.

\section{Icosahedral multi-component model sets}

We shall re-define the pattern ${\mathcal Q}$ as a multi-component model set 
by using a 6-dimensional subspace of $\mathbb{E}_k$.
The automorphism 
\begin{equation}
\varphi :\mathbb{Q}[\tau ]\longrightarrow \mathbb{Q}[\tau ]
\end{equation}
of the quadratic field $\mathbb{Q}[\tau ]$ that maps $\sqrt{5}\mapsto -\sqrt{5}$
has the property $\varphi (\tau )=\tau '$. The representation (\ref{Y}) is related
through $\varphi $ to the representation
$\{ T'_g:\mathbb{E}_3\longrightarrow \mathbb{E}_3\ |\ g\in Y\}$ 
belonging to $\Gamma _3$
generated by the rotations $T'_a,\ T'_b :\mathbb{E}_3\longrightarrow \mathbb{E}_3$
\begin{equation}\label{Y'} \fl \begin{array}{l}
T'_a(\alpha ,\beta  ,\gamma )=
\left(\frac{\tau '-1}{2}\alpha -\frac{\tau '}{2}\beta  +\frac{1}{2}\gamma ,
\ \frac{\tau '}{2}\alpha +\frac{1}{2}\beta  +\frac{\tau '-1}{2}\gamma ,
\ -\frac{1}{2}\alpha +\frac{\tau '-1}{2}\beta  
+\frac{\tau '}{2}\gamma \right)\\[1mm]
T'_b(\alpha ,\beta  ,\gamma )=(-\alpha ,-\beta  ,\gamma ).
\end{array} \end{equation}
If instead of the representation (\ref{Y}) and cluster ${\mathcal C}$ we start from 
the representation (\ref{Y'}) and the cluster
\begin{equation}
{\mathcal C}'=\{ e'_1,e'_2,...,e'_k,-e'_1,-e'_2,...,-e'_k\}
\end{equation}
where 
\begin{equation}
e'_i=(e'_{i1},e'_{i2},e'_{i3})=(\varphi (e_{i1}),\varphi (e_{i2}),\varphi (e_{i3}))
\end{equation}
then we get the same representation of $Y$ in $\mathbb{E}_k$ and the $Y$-invariant
subspace
\begin{equation}
E'=\left\{ \ (<u,e'_1>,<u,e'_2>,...,<u,e'_k>)\ |\ \  u\in \mathbb{E}_3\ \right\} .
\end{equation}

The vectors 
\[ \fl v'_1=\varrho '(e'_{11},e'_{21},...,e'_{k1})\qquad 
   v'_2=\varrho '(e'_{12},e'_{22},...,e'_{k2})\qquad 
   v'_3=\varrho '(e'_{13},e'_{23},...,e'_{k3})\] 
where $\varrho '=1/\sqrt{(e'_{11})^2+(e'_{21})^2+...+(e'_{k1})^2}$,
form an orthonormal basis of $E'$, and the matrix of the orthogonal projector
 \ $\pi ':\mathbb{E}_k\longrightarrow \mathbb{E}_k$ corresponding to $E'$ in the basis
$\{ \varepsilon _1,\varepsilon _2,...,\varepsilon _k\}$  is
\begin{equation}
 \pi '={\varrho '}^2\left( \begin{array}{llll}
\langle e'_1,e'_1\rangle & \langle e'_1,e'_2\rangle & ... & \langle e'_1,e'_k\rangle \\
\langle e'_2,e'_1\rangle & \langle e'_2,e'_2\rangle & ... & \langle e'_2,e'_k\rangle \\
...& ... & ... & ...\\
\langle e'_k,e'_1\rangle & \langle e'_k,e'_2\rangle & ... & \langle e'_k,e'_k\rangle 
\end{array}  \right) .
\end{equation}
{\bf Theorem 6.} {\it The projectors $\pi $ and $\pi '$ are orthogonal, that is, 
\[ \pi \pi '=\pi '\pi =0 \]
and the projector $\pi +\pi '$ corresponding to the subspace
${\mathcal E}=E\oplus E'$ has rational entries.}\\[2mm]
{\bf Proof}. Consider the linear mapping 
\[ A:\mathbb{E}_3\longrightarrow \mathbb{E}_3:\, u\mapsto Au\qquad 
{\rm where}\quad Au=\sum_{i=1}^k\langle u,e_i\rangle e'_i .\]
Since $A$ is a morphism of representations
\[ A(T_gu)=\sum_{i=1}^k\langle T_gu,e_i\rangle e'_i
=\sum_{i=1}^k\langle u,T_g^{-1}e_i\rangle e'_i\]
\[ =T'_g\left( \sum_{i=1}^k\langle u,T_g^{-1}e_i\rangle {T'}_g^{-1}e'_i\right)
=T'_g\left( \sum_{i=1}^k\langle u,e_i\rangle e'_i\right)=T_g(Au)\]
between the irreducible non-equivalent representations (\ref{Y}) and (\ref{Y'}),
from Schur's lemma it follows that $A=0$, that is, 
$\sum_{i=1}^k\langle u,e_i\rangle e'_i=0$ for any $u\in \mathbb{E}_3$.
Particularly, we have
\[ \sum_{i=1}^k\langle e_j,e_i\rangle \langle e'_i,e'_l\rangle =
\langle \sum_{i=1}^k\langle e_j,e_i\rangle e'_i,e'_l\rangle =0 \]
whence $\pi \pi '=0$. In a similar way we can prove that $\pi '\pi =0$.
Since 
\[ {\varrho '}^2\langle e'_i,e'_j\rangle 
=\varphi \left( \varrho ^2\langle e_i,e_j\rangle \right) \]
we get ${\varrho '}^2\langle e'_i,e'_j\rangle 
+\varrho ^2\langle e_i,e_j\rangle \in \mathbb{Q}$, that is, the projector
$\pi +\pi '$ has rational entries. \\[3mm]
{\bf Theorem 7.} {\it  The collection of spaces and mappings
\begin{equation}\begin{array}{ccccccc}
\pi x \leftarrow x 
&:E& \stackrel{\pi }\longleftarrow & {\mathcal E}
& \stackrel{\pi ' }\longrightarrow & E' :& x \rightarrow \pi ' x\\
&&&\cup &&& \\
&&& {\mathcal L} &&&
\end{array}
\end{equation} 
where ${\mathcal L}=(\pi +\pi ')(\mathbb{L})$, 
is a  cut and project scheme} \cite{B,M}.\\[2mm]
{\bf Proof.} Since, in view of theorem 5, we have 
\[ \pi '({\mathcal L})=\pi '(\pi +\pi ')(\mathbb{L})=\pi '(\mathbb{L})=
\sum_{i=1}^k\mathbb{Z}e'_i \]
the set $\pi '({\mathcal L})$ is dense in $E'$.
For each $x\in {\mathcal L}$ there is $\kappa y\in \mathbb{L}$ with 
$y\in \mathbb{Z}^k$ such that $x=(\pi +\pi ')(\kappa y)$. If $\pi x=0$
then $\pi (\pi +\pi ')(\kappa y)=0$, whence $\pi (\kappa y)=0$.
But, $\pi (\kappa y)=\kappa \pi y$, and hence we have $\pi y=0$.
Since $y\in \mathbb{Z}^k$, from $\pi y=0$ we get $\pi 'y=0$, whence
$x=(\pi +\pi ')y=0$. This means that $\pi $ restricted to ${\mathcal L}$
is injective.\\

Let $E''={\mathcal E}^\perp =
\{ x\in \mathbb{E}_k\ |\ \langle x,y\rangle =0\  
{\rm for\ all\ } y\in {\mathcal E}\}$ and let 
$\pi '' :\mathbb{E}_k\longrightarrow \mathbb{E}_k$, $\pi ''x=x-\pi x-\pi 'x$
be the corresponding orthogonal projector.
The lattice $L=\mathbb{L}\cap {\mathcal E}$ is a sublattice of ${\mathcal L}$,
and necessarily $[{\mathcal L}:L]$ is finite. Since $\pi ''$ has rational 
entries the projection $\mathbb{L}''=\pi ''(\mathbb{L})$ of 
$\mathbb{L}$ on $E''$ is a discrete countable set. 
Let ${\mathcal Z}=\{ z_i\ |\ i\in \mathbb{Z} \}$ be a subset of $\mathbb{L}$ such
that $\mathbb{L}''=\pi ''({\mathcal Z})$ and $\pi ''z_i\not=\pi ''z_j$ for $i\not=j$. 
The lattice $\mathbb{L}$ is contained in the union of the cosets
${\mathcal E}_i=z_i+{\mathcal E}=\{ z_i+x\ |\  x\in {\mathcal E}\}$ 
\begin{equation}
\mathbb{L}\subset \bigcup_{i\in\mathbb{Z}} {\mathcal E}_i.
\end{equation}
Since $\mathbb{L}\cap {\mathcal E}_i=z_i+L$ the set
\begin{equation}
{\mathcal L}_i=(\pi +\pi ')(\mathbb{L}\cap {\mathcal E}_i)=
(\pi +\pi ')z_i+L
\end{equation}
is a coset of $L$ in ${\mathcal L}$ for any $i\in \mathbb{Z}$.

Only for a finite number of cosets ${\mathcal E}_i$ the intersection 
\begin{equation}
K_i=K\cap {\mathcal E}_i=\pi ^\perp (\mathbb{K}\cap {\mathcal E}_i)
\subset \pi ''z_i+E'
\end{equation}
is non-empty.
By changing the indexation of the elements of ${\mathcal Z}$
if necessary, we can assume that the subset of $E'$
\begin{equation}
{\mathcal K}_i=\pi '(K_i)=\pi '(\mathbb{K}\cap {\mathcal E}_i)\subset E'
\end{equation}
has a non-empty interior only for $i\in \{ 1,...,m\}.$ 
The `polyhedral' set ${\mathcal K}_i$ satisfies the conditions:

\begin{itemize}
\item[(a)] ${\mathcal K}_i\subset E'$ is compact;
\item[(b)] ${\mathcal K}_i=\overline{{\rm int}({\mathcal K}_i)}$;
\item[(c)] The boundary of ${\mathcal K}_i$ has Lebesgue measure $0$
\end{itemize}
for any $i\in \{ 1,...,m\}.$
This allows us to re-define ${\mathcal Q}$ in terms of the 6-dimensional
superspace ${\mathcal E}$ as a multi-component model set \cite{B}
\begin{equation}
{\mathcal Q} =\bigcup_{i=1}^m
\left\{ \pi x\ \left| \ x\in {\mathcal L}_i , \ \pi ' x\in {\mathcal K}_i
\right.\right\}.
\end{equation} 
It is known \cite{D} that this is the minimal embedding
for a 3-dimensional quasiperiodic point set with icosahedral symmetry.
The main difficulty in this new approach is the determination of the 
`atomic surfaces' ${\mathcal K}_i$.

\ack{This research was supported by the grant CNCSIS no. 630/2003.}

\end{document}